\documentclass[12pt]{iopart}

\usepackage{iopams}
\usepackage{graphicx}
\usepackage{pseudocode}
 \expandafter\let\csname equation*\endcsname\relax
  \expandafter\let\csname endequation*\endcsname\relax
\usepackage{amsmath}
\usepackage{amssymb}
\usepackage{amsthm}  
\usepackage{color}
\newcommand{\vm}[1]{\boldsymbol{#1}}
\renewcommand{\thepseudocode}{1} 

\begin{document}

\title{Statistical mechanics analysis of thresholding 1-bit compressed sensing}

\author{Yingying~Xu$^{1}$ and Yoshiyuki~Kabashima$^{2}$}

\address{$^{1}$Department of Computer Science, School of Science, 
\\Aalto University,
P.O.Box 15400, FI-00076 Aalto, FINLAND

$^{2}$Department of Mathematical and Computing Science, 
\\Tokyo Institute of Technology, 
\\G5-22, 4259 Nagatsuda-chou, Midori-ku, Yokohama, Kanagawa, 226-8502, JAPAN}
\ead{yingying.xu@aalto.fi, kaba@c.titech.ac.jp}

\begin{abstract}
The one-bit compressed sensing (1bit CS) framework aims to reconstruct a sparse signal by only using the sign information of its linear measurements. To compensate for the loss of scale information, past studies in the area have proposed recovering the signal by imposing an additional constraint on the $l_2$-norm of the signal. Recently, an alternative strategy that captures scale information by introducing a threshold parameter to the quantization process was advanced. In this paper, we analyze the typical behavior of the thresholding 1-bit compressed sensing utilizing the replica method of statistical mechanics, so as to gain an insight for properly setting the threshold value. Our result shows that, fixing the threshold at a constant value yields better performance than varying it randomly when the constant is optimally tuned, statistically. Unfortunately, the optimal threshold value depends on the statistical properties of the target signal, which may not be known in advance. In order to handle this inconvenience, we develop a heuristic that adaptively tunes the threshold parameter based on the frequency of positive (or negative) values in the binary outputs. Numerical experiments show that the heuristic exhibits satisfactory performance while incurring low computational cost. 
\end{abstract}

\maketitle

\section{Introduction}
For the last decade, {\em compressed sensing} (CS) has received considerable attention as a novel technology in signal processing research. The purpose of CS is to enhance signal processing performance by utilizing the notion of the {\em sparsity} of signals \cite{CScandes}--\cite{StarckMurtaghFadili2010}. Let us suppose that a sparse vector $\vm{x^{0}} \in\mathbb{R}^{N}$, many components of which are zero, is linearly transformed into vector $\vm{y} \in\mathbb{R}^{M}$ by an $M \times N$ measurement matrix $\vm{\Phi}$, where $\vm{y} = \vm{\Phi x^{0}}$. For a given pair of $\vm{\Phi}$ and $\vm{y}$, the reconstruction of $\vm{x^{0}}$ is required \cite{CandesWakin2008}. Many studies in CS research have shown that the sparsity of signals makes it possible to perfectly reconstruct  $\vm{x}^0$ at a viable computational cost, even in the region of $\alpha = M/N <1$ \cite{CandesRombergTao2006}--\cite{Krzakala2012}. This has led to the hardware-level realization of accurate signal reconstruction that had hitherto been regarded as out of reach due to limitations on sampling rates \cite{bandlimit} and/or the number of sensors \cite{singlepixel}.  

In the signal processing context, the CS framework eases the burden on analog-to-digital converters (ADCs) by reducing the sampling rate required to acquire and recover sparse signals. However, in practice, ADCs not only sample, but also quantize each measurement to a finite number of bits; moreover, there is an inverse relationship between achievable sampling rate and bit depth. Therefore, many discussions on CS have shifted emphasis from sampling rate to number of bits per measurement \cite{measurementsVSbits, ratedistortion}. In particular, we are here interested in the extreme case of 1-bit CS measurement, which captures just the sign as $\vm{y} = \text{sign}(\vm{\Phi x^{0}})$ \cite{1bitCSbaranuik}. Thus, the measurement operator is a mapping from $\mathbb{R}^{N}$ to the Boolean cube $B^{M}:= \{ -1, 1 \} ^M$. This is highly beneficial in practice due to the significant reduction in the size of data that are transmitted and stored.

It is obvious that the scale (absolute amplitude) of the signal is lost in 1-bit CS measurements. To compensate for this, past studies have proposed the imposition of an additional constraint whereby the $l_2$-norm of the signal is normalized to a fixed constant \cite{1bitCSbaranuik, 1bitCSL1XuKaba}. In other words, this can only reconstruct the directional information but not the true scale information of the signal. Moreover, it yields another drawback such that solving the reconstruction problem becomes nontrivial, since the problem is no longer formulated as a convex optimization. In order to address these issues, by introducing a set of finite thresholds $\vm{\lambda} = (\lambda_\mu)$ $(\mu = 1, 2, \ldots, M)$ to the quantizer as

\begin{equation}
\vm{y} = \text{sign}(\vm{\Phi x} + \vm{\lambda}), 
\label{1-bit-measurement}
\end{equation}
and combining the knowledge of the thresholds,  we are able to estimate the scale of the signal {{\cite{1bitCSNormEstKKR}}}. 
Furthermore, as the feasible set provided by the constraint of (\ref{1-bit-measurement}) for given measurements $\vm{y}$ is a convex region of $\vm{x}$,  one can reconstruct a sparse signal in polynomial time by solving the $l_1$-norm minimization problem 
\begin{equation}
  \vm{\hat{ x}} = \underset{\vm{x} \in\mathbb{R}^{N}}{\text{argmin }} ||\vm{x}||_{1} \text{ subject to }  \vm{y} = \text{sign}(\vm{\Phi x} + \vm{\lambda}) 
\label{thresh1bitCS}  
\end{equation}
by using versatile convex optimization algorithms \cite{cvx}. 

A lingering, natural question is how should we set the values of $\lambda_\mu$. To partially answer this, we compare two strategies: one involves fixing the thresholds at a constant value $\lambda_\mu = \lambda$ for all measurements, and the other consists of independently selecting $\lambda_\mu$ from an identical Gaussian distribution. 
{{In \cite{1bitCSNormEstKKR}, worst case bounds of the number of measurements necessary for achieving permissible reconstruction errors are evaluated for the two strategies. However, worst case evaluations, in general, do not necessarily well describe the performance actually observed in practical situations, and therefore, alternative investigations for probing the typical performance are also important. Having this perspective, we here analyze the typical performance of the thresholding 1-bit CS using statistical mechancis.}}
We will show that the fixing-value strategy statistically yields better mean squared error (MSE) performance than the random strategy when adjustable parameters are optimally tuned using the replica method \cite{replica} of statistical mechanics.

Unfortunately, the value of the optimal threshold depends on the statistical property of the target signal, which may not be known in advance. To cope with such situations, we focus here on the distribution of binary output y, which indirectly conveys the {amplitude information of the target signal $\vm{x^0}$} and can be estimated from measurements. We develop a heuristic that adaptively tunes the threshold parameter based on the frequency of positive (or negative) values in the binary outputs.
 Numerical experiments show that our algorithm exhibits satisfactory {performance which is comparable to that} achieved by the optimally tuned threshold. 

The rest of this paper is organized as follows: In Section II, we formulate the problem to be addressed in this study. In Section III, we evaluate the performance of the reconstruction method of (\ref{thresh1bitCS}). Section IV is devoted to a description of our learning algorithm to tune the threshold value, whereas Section V summarizes our work in this study.

\section{Problem set up}
Let us suppose a situation where entry $x_i^0$ $(i = 1, 2, \ldots, N)$ of $N$-dimensional signal $\vm{x}^0 \in \mathbb{R}^N$ is independently generated from an identical sparse distribution:
\begin{equation}
P\left(x\right) = \left(1 - \rho\right) \delta \left( x \right)
+ \rho\tilde{P} \left( x \right), 
\label{sparse}
\end{equation}
where $\rho\in[0, 1]$ represents the density of nonzero entries in the signal, and $\tilde{P} (x)$ is a distribution function of $x \in \mathbb{R}$ that does not have finite mass at $x = 0$. In the thresholding 1-bit CS, the measurement is performed as 
\begin{equation}
\textrm{\boldmath $y$} = \mathrm{sign} \left(\vm{\Phi x^{0}} + \vm{\lambda} \right), 
\label{measurement}
\end{equation}
where we assume that each entry of the $M \times N$ measurement matrix $\vm{\Phi}$ is provided as an independent sample from a Gaussian distribution of mean zero and variance $N^{-1}$.

We consider two strategies for setting the thresholding vector $\vm{\lambda} = (\lambda_\mu)$. Case 1: entry $\lambda_\mu = \lambda$ is fixed for all $\mu = 1, 2, \ldots, M$. Case 2: $\lambda_\mu$ is independently sampled from a Gaussian distribution ${\cal N}(0, \sigma_{\lambda}^2)$. For both cases, the feasible set consistent with given outputs $\vm{y}$ is provided by a set of inequalities
\begin{equation}
 y_\mu \left (\sum_{i = 1}^N \Phi_{\mu i} x_i + \lambda_\mu  \right ) > 0
 \label{feasible_region}
 \end{equation}
 $(\mu = 1, 2, \ldots, M)$, which defines a convex region of $\vm{x}$. Therefore, a sparse signal is reconstructed by the $l_1$-norm minimization (\ref{thresh1bitCS}) utilizing a certain convex optimization algorithm.

\section{Analysis}
\subsection{Method}
The key to finding the statistical properties of reconstruction (\ref{thresh1bitCS}) is the average free energy density
\begin{equation}
 \bar{f} \equiv - \lim_{\beta, N \to \infty}\frac{1}{\beta N} \left [\ln Z(\beta; \vm{\Phi}, \vm{x}^0, \vm{\lambda}) \right]_{\vm{\Phi},\vm{x}^0, \vm{\lambda}},
\end{equation}
where
\begin{equation}
Z\left(\beta;\vm{\Phi},\vm{x^{0}},\vm{\lambda}\right)
\!=\!\int d \textrm{\boldmath $x$}e^{-\beta\vert\vert\textrm{\boldmath $x$}\vert\vert _{1}}
\prod_{\mu=1}^M \Theta\left ( (\vm{\Phi} \vm{x}^0 +\vm{\lambda})_\mu  (\vm{\Phi} \vm{x}+\vm{\lambda})_\mu \right )
\label{eq:partition_func}
\end{equation}
is the partition function. {{We consider the large system size limit, $N\rightarrow\infty ,M\rightarrow\infty$, while keeping $\alpha=M/N$ finite.}} Here, $\Theta\left(x \right) = 1$ and $0$ for $x > 0$ and $x < 0$, respectively, offers the basis for our analysis. 
$\left [\cdots \right]_{X}$ generally denotes the operation of the average with respect to the random variable $X$.
As $\beta$ tends to infinity, the integral of (\ref{eq:partition_func}) is dominated by the correct solution of (\ref{thresh1bitCS}), {which offers the minimum $l_1$-norm of $\vm{x}$}. One can therefore evaluate the performance of the solution by examining the macroscopic behavior of (\ref{eq:partition_func}) in the limit of $\beta\rightarrow \infty$. Because directly averaging the logarithm of the partition function is difficult, we employ the replica method \cite{replica}, which allows us to calculate the average free energy density as 
\begin{equation}
 \bar{f} = -\lim_{n \to +0}\frac{\partial}{\partial n}\lim_{\beta, N\to\infty}\frac{1}{\beta N} \ln \left [Z^{n}(\beta; \vm{\Phi}, \vm{x}^0,\vm{\lambda}) \right]_{\vm{\Phi}, \vm{x}^0,\vm{\lambda}}.
\end{equation}

For this, we first evaluate the $n$-th moment of the partition function $\left [Z^n \left(\beta; \vm{\Phi}, \textrm{\boldmath $x$}^{0},\vm{\lambda} \right) \right]_{\vm{\Phi}, \vm{x}^0, \vm{\lambda}}$ for $n = 1, 2, \ldots \in \mathbb{N}$ by using the formula 
\begin{equation}
Z^n \left(\beta; \vm{\Phi}, \textrm{\boldmath $x$}^{0},\vm{\lambda} \right)
=
\int \prod_{a=1}^n \left (d \textrm{\boldmath $x$}^a
e^{-\beta\vert\vert\textrm{\boldmath $x^a$}\vert\vert _{1}}  \right )
 \times 
\prod_{a=1}^n \prod_{\mu=1}^M \Theta\left ( (\vm{\Phi} \vm{x}^0 +\vm{\lambda})_\mu  (\vm{\Phi} \vm{x}^a+\vm{\lambda})_\mu \right ), 
\label{eq:expansion}
\end{equation}
which holds only for $n = 1, 2, \ldots \in \mathbb{N}$. Here, $\vm{x}^a$ ($a = 1, 2, \ldots, n$) denotes the $a$-th replicated signal. Averaging (\ref{eq:expansion}) with respect to $\vm{\Phi}$ and $\vm{x}^0$ results in the saddle point evaluation concerning macroscopic variables $q_{0a} = q_{a0}\equiv N^{-1} \vm{x}^0 \cdot \vm{x}^a$ and $q_{ab} = q_{ba} \equiv N^{-1} \vm{x}^a \cdot \vm{x}^b$ ($a, b = 1, 2, \ldots, n$). Although (\ref{eq:expansion}) holds only for $n \in \mathbb{N}$, the expression $(\beta N)^{-1} \ln \left [Z^n \left(\beta; \vm{\Phi}, \textrm{\boldmath $x$}^{0},\vm{\lambda} \right) \right]_{\vm{\Phi}, \vm{x}^0, \vm{\lambda}}$ obtained by the saddle point evaluation, under a certain assumption concerning the permutation symmetry with respect to the replica indices $a, b$, is obtained as an analytic function of $n$, which is likely to also hold for $n \in \mathbb{R}$. Therefore, we utilize the analytic function to evaluate the average of the logarithm of the partition function to obtain $\bar{f}$. 

In particular, under the replica symmetric (RS) ansatz, where the dominant saddle point is assumed to be of the form 
\begin{eqnarray}
q_{ab}=q_{ba}=\left \{
\begin{array}{ll}
Q_{0} & (a=b=0) \cr
m & (a=1,2,\ldots,n; \ b=0) \cr
Q & (a=b=1,2,\ldots,n) \cr
q & (a\ne b =1,2,\ldots,n) 
\end{array}
\right . . 
\label{RSanzats}
\end{eqnarray}
{{The problem setting so far is applicable generally for any distribution $\tilde{P}(x)$ in (\ref{sparse}).}} For simplicity, we hereafter assume that $x^0$ is distributed from (\ref{sparse}) with $\tilde{P}(x) = {\cal N} (0, \sigma_{0}^2)$; therefore $Q_{0} = \rho\sigma_{0}^{2}$.

\subsection{Resulting equations}
The above procedure (\ref{RSanzats}) offers an expression of the average free-energy density as 
\begin{eqnarray}
\bar{f} 
  &=& \mathop{\rm extr}_{\omega} \Biggr\{\! \int\! {\rm D}z P(x^{0})\phi \left(\sqrt{\hat{q}}z+\hat{m} x^{0};\hat{Q}\right)
-\frac{1}{2}\hat{Q}q+\frac{1}{2}\hat{q}\chi+\hat{m}m {\sigma_{0}^2}\nonumber\\
 &&+\frac{\alpha}{2\chi}\left[{H}\left (-\frac{\frac{m}{\sqrt{q}}t+\lambda}{\sqrt{\rho{\sigma_{0}^2}-\frac{m^2}{q}}} \right )\left(\sqrt{q}t+\lambda\right)^2\Theta\left(-\sqrt{q}t-\lambda\right)\right.\nonumber\\
 &&\left.+{H}\left (\frac{\frac{m}{\sqrt{q}}t+\lambda}{\sqrt{\rho{\sigma_{0}^2}-\frac{m^2}{q}}} \right )\left(\sqrt{q}t+\lambda\right)^2\Theta\left(\sqrt{q}t\!+\!\lambda\!\right)\!\right]_{t,\lambda} \!\Biggr\}\! 
\label{eq:free energy}
\end{eqnarray}
in the limit of $\beta \to \infty$. Here, $\alpha = M/N$, $\textrm{extr}_{X}\{g(X)\}$ denotes the extremization of function $g(X)$ with respect to $X$, $\omega = \{\chi, m, q, \hat{Q}, \hat{q}, \hat{m}\}$, ${H}(x) = \int_x^{+\infty} {\rm D}z$, $\textrm{D}z = \textrm{d}z \textrm{exp}(-z^2/2)/\sqrt{2\pi}$ is a Gaussian measure, $t$ and $z$ are independent and identically distributed (i.i.d) random variables from ${\cal N}(0, 1)$. The function $\phi(h; \hat{Q})$ is defined as
\begin{eqnarray}
\phi(h;\hat{Q})&=&\mathop{\rm min}_{x} \left \{ 
\frac{\hat{Q}}{2} x^2-h x + |x| \right \}
= -\frac{1}{2\hat{Q}} \left (|h|-1 \right )^2 \Theta\left ( |h| -1 \right ).
\end{eqnarray}
The derivation of $(\ref{eq:free energy})$ is provided in \ref{replicaderivation}.

For Case 1, which fixes the threshold for all measurements to a constant $\lambda$ as $\lambda_\mu = \lambda$ $(\mu = 1, 2, \ldots, M)$, the extremization of (\ref{eq:free energy}) is reduced to the following saddle point equations: 
\begin{flushleft}
\begin{eqnarray}
\hat{q}&\!=\!&\frac{\alpha}{\chi^2}\left\{\left[ {H}\left(-\frac{\frac{mt}{\sqrt{q}}+\lambda}{\sqrt{\rho\sigma_{0}^2-\frac{m^2}{q}}}\right)u\left(-\sqrt{q}t-\lambda\right)\right.\right]_t \nonumber\\
&&\left.+\left[{H}\left(\frac{\frac{mt}{\sqrt{q}}+\lambda}{\sqrt{\rho\sigma_{0}^2-\frac{m^2}{q}}}\right)u\left(\sqrt{q}t+\lambda\right)\right]_{t}\right\},\label{qh_fix}\\
\hat{Q}&\!=\!&\frac{\alpha}{\chi} \left\{ \left[{H}\left(-\frac{\frac{mt}{\sqrt{q}}+\lambda}{\sqrt{\rho\sigma_{0}^2-\frac{m^2}{q}}}\right)u''\left(-\sqrt{q}t-\lambda\right)\right]_t\right.\nonumber\\
&&\left.+\left[{H}\left(\frac{\frac{mt}{\sqrt{q}}+\lambda}{\sqrt{\rho\sigma_{0}^2-\frac{m^2}{q}}}\right)u''\left(\sqrt{q}t+\lambda\right)\right]_t\right\},\label{Qh_fix}\\
\hat{m}&\!=\!&\frac{\alpha}{\chi\sqrt{2\pi\left(\rho\sigma_{0}^2-\frac{m^2}{q}\right)}}\left[\text{exp}\left(-\frac{\left(\frac{mt}{\sqrt{q}}+\lambda\right)^2}{2\left(\rho\sigma_{0}^2-\frac{m^2}{q}\right)}\right)\right.\nonumber\\ 
&&\left.\times\left(u'\left(\sqrt{q}t+\lambda\right)-u'\left(-\sqrt{q}t-\lambda\right)\right)\right]_t, \label{mh_fix}\\
q&\!=\!&\frac{2}{\hat{Q}^{2}}\left\{ \left(1-\rho\right)\left(\left(\hat{q}+1\right){H}\left(\frac{1}{\sqrt{\hat{q}}}\right)
          -\sqrt{\frac{\hat{q}}{2\pi}}e^{-\frac{1}{2\hat{q}}} \right) \right . \nonumber\\
&&\left. +\rho\left(\left(\hat{q}+\hat{m}^2\sigma_{0}^2+1\right){H}\left(\frac{1}{\sqrt{\hat{q}+\hat{m}^2\sigma_{0}^2}}\right)\right.\right.\nonumber\\
&&\left.\left.-\sqrt{\frac{\hat{q}+\hat{m}^2\sigma_{0}^2}{2\pi}}e^{-\frac{1}{2\left(\hat{q}+\hat{m}^2\sigma_{0}^2\right)}}\right)\right\}, \label{q_fix}\\
\chi &\!=\!&\frac{2}{\hat{Q}}\left\{\left(1-\rho\right)
{H}\left(\frac{1}{\sqrt{\hat{q}}}\right)\!+\!\rho {H}\!\left(\!\frac{1}{\sqrt{\hat{q}+\hat{m}^2\sigma_{0}^2}}\!\right)\!\right\}\!,\! \label{chi_fix}\\
m&\!=\!&\frac{2\rho\hat{m}\sigma_{0}^2}{\hat{Q}}{H}\left(\frac{1}{\sqrt{\hat{q}+\hat{m}^2\sigma_{0}^2}}\right)\label{m_fix}, 
\end{eqnarray}
\end{flushleft}
where $u(x) = x^2\Theta(x)$, and $t$ obeys the standard normal distribution ${\cal N}(0, 1)$. 

On the other hand, for Case 2, where $\lambda_\mu$ is sampled independently from ${\cal N}(0, \sigma^2_\lambda)$ for $\mu = 1, 2, \ldots, M$, the saddle point equations of $\hat{q}$, $\hat{Q}$, and $\hat{m}$ are modified to 
\begin{eqnarray}
\hat{q}&\!=\!&\frac{2\alpha}{\chi^2}\left[{H}\left(\frac{\frac{mt}{\sqrt{q}}+\sigma_{\lambda}r}{\sqrt{\rho\sigma_{0}^2-\frac{m^2}{q}}}\right)u\left(\sqrt{q}t+\sigma_{\lambda}r\right)\right]_{r,t},\label{qh_Gauss}\\
\hat{Q}&\!=\!&\frac{2\alpha}{\chi}\left[{H}\left(\frac{\frac{mt}{\sqrt{q}}+\sigma_{\lambda}r}{\sqrt{\rho\sigma_{0}^2-\frac{m^2}{q}}}\right)u''\left(\sqrt{q}t+\sigma_{\lambda}r\right)\right]_{r,t},\label{Qh_Gauss}\\
\hat{m}&\!=\!&\frac{2\alpha}{\chi\sqrt{2\pi\left(\rho\sigma_{0}^2-\frac{m^2}{q}\right)}}\left[\text{exp}\left(-\frac{\left(\frac{mt}{\sqrt{q}}+\sigma_{\lambda}r\right)^2}{2\left(\rho\sigma_{0}^2-\frac{m^2}{q}\right)}\right)\right.\nonumber\\
&&\times\left.u'\left(\sqrt{q}t+\sigma_{\lambda}r\right)\right]_{r,t},\label{mh_Gauss}
\end{eqnarray}
where $r$ is a variable sampled from the standard normal distribution ${\cal N}(0, 1)$. The remaining equations for $q$, $\chi$, and $m$ are identical to (\ref{q_fix}), (\ref{chi_fix}), and (\ref{m_fix}), respectively.

\subsection{Simulations and observations}
The value of $m$ determined by these equations physically represents the typical overlap $N^{-1} \left [\vm{x}^0 \cdot \hat{\vm{x}} \right]_{\vm{\Phi}, \vm{x}^0, \vm{\lambda}} $ between the original signal $\vm{x}^0$ and the solution $\hat{\vm{x}}$ of (\ref{thresh1bitCS}). Therefore, the typical value of MSE between $\vm{x}^0$ and $\hat{\vm{x}}$, which serves as the performance measure of the reconstruction problem, is evaluated as 
\begin{eqnarray}
{\rm MSE} = N^{-1}\left [ \left |\hat{\vm{x}}-\vm{x}^0\right |^2 \right ]_{\vm{\Phi,\vm{x}^0},\vm{\lambda}}
=q+ \rho\sigma_{0}^2-2m.
\label{MSE}
\end{eqnarray}
Note that in past studies on 1-bit CS, reconstruction performance was evaluated through directional MSE, which is defined by $|\frac{\hat{\vm{x}}}{|\hat{\vm{x}}|} - \frac{\vm{x}^0}{|\vm{x}^0|}|^2$ as scale information is lost. 

\begin{figure}[h]
\begin{center}
  \includegraphics[width=3.5in]{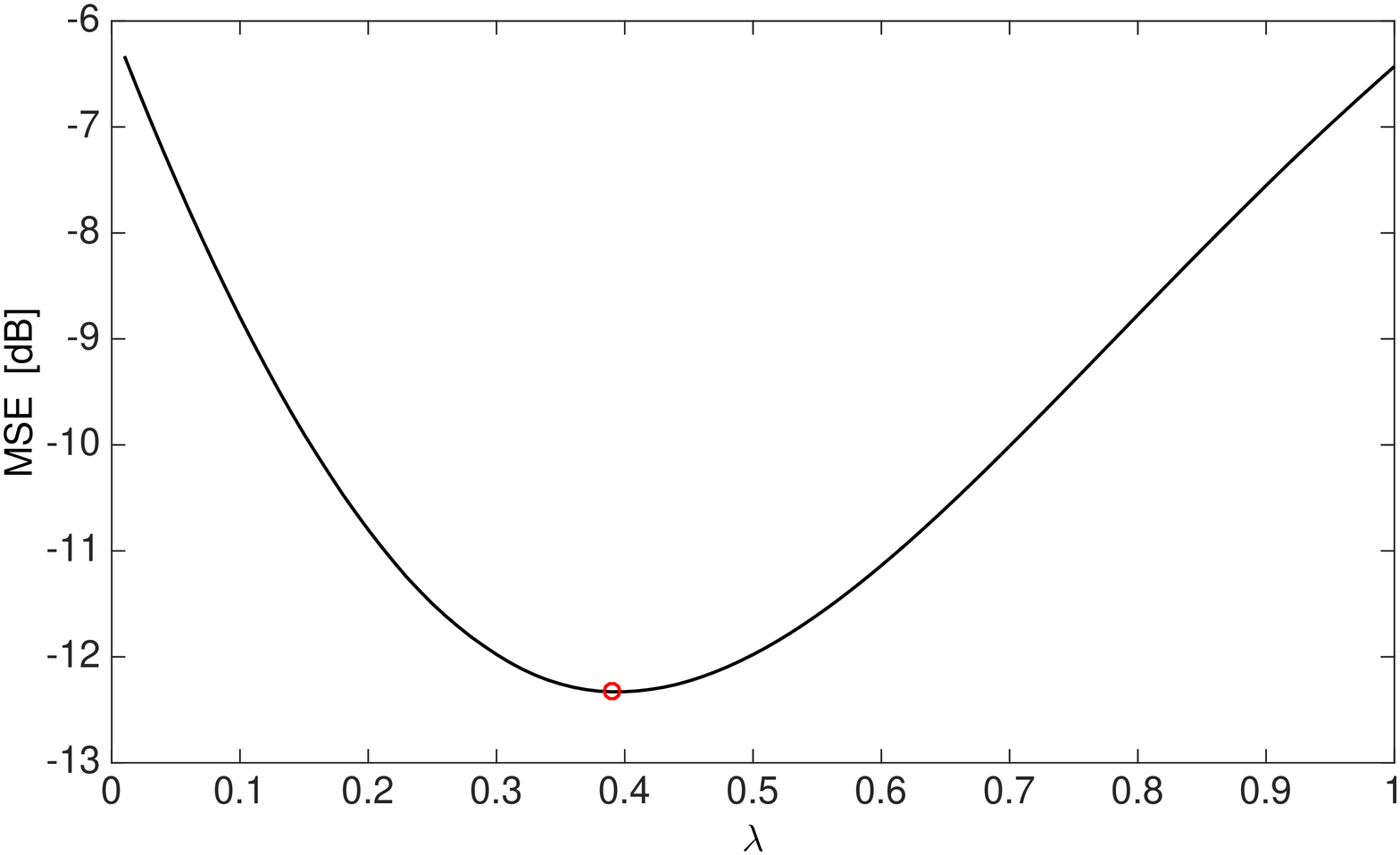}
  \caption{Replica prediction of MSE (in decibel) versus fixed threshold $\lambda$ for signal distribution $\rho = 0.25, \sigma_{0}^{2} = 1$, and ratio $\alpha = 3$.}
  \label{fig:MSEdb_fix}
\end{center}
\end{figure}

\begin{figure}[h]
\begin{center}
  \includegraphics[width=3.5in]{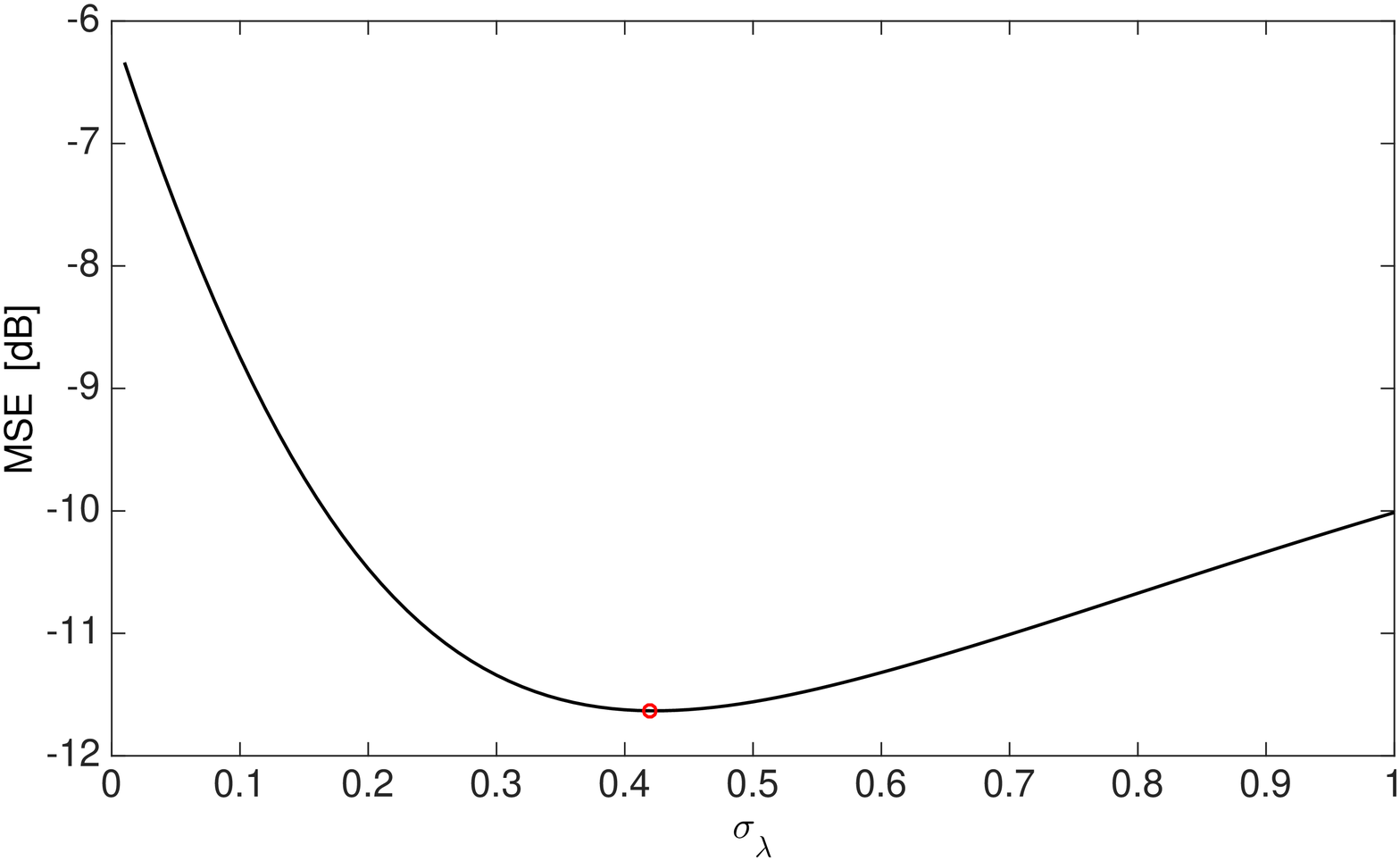}
  \caption{Replica prediction of MSE (in decibel) versus $\sigma_{\lambda}$ for signal $\rho = 0.25, \sigma_{0}^{2} = 1$, and ratio $\alpha = 3$.}
  \label{fig:MSEdb_Gauss}
\end{center}
\end{figure}

\begin{figure}[h]
\begin{center}
  \includegraphics[width=3.5in]{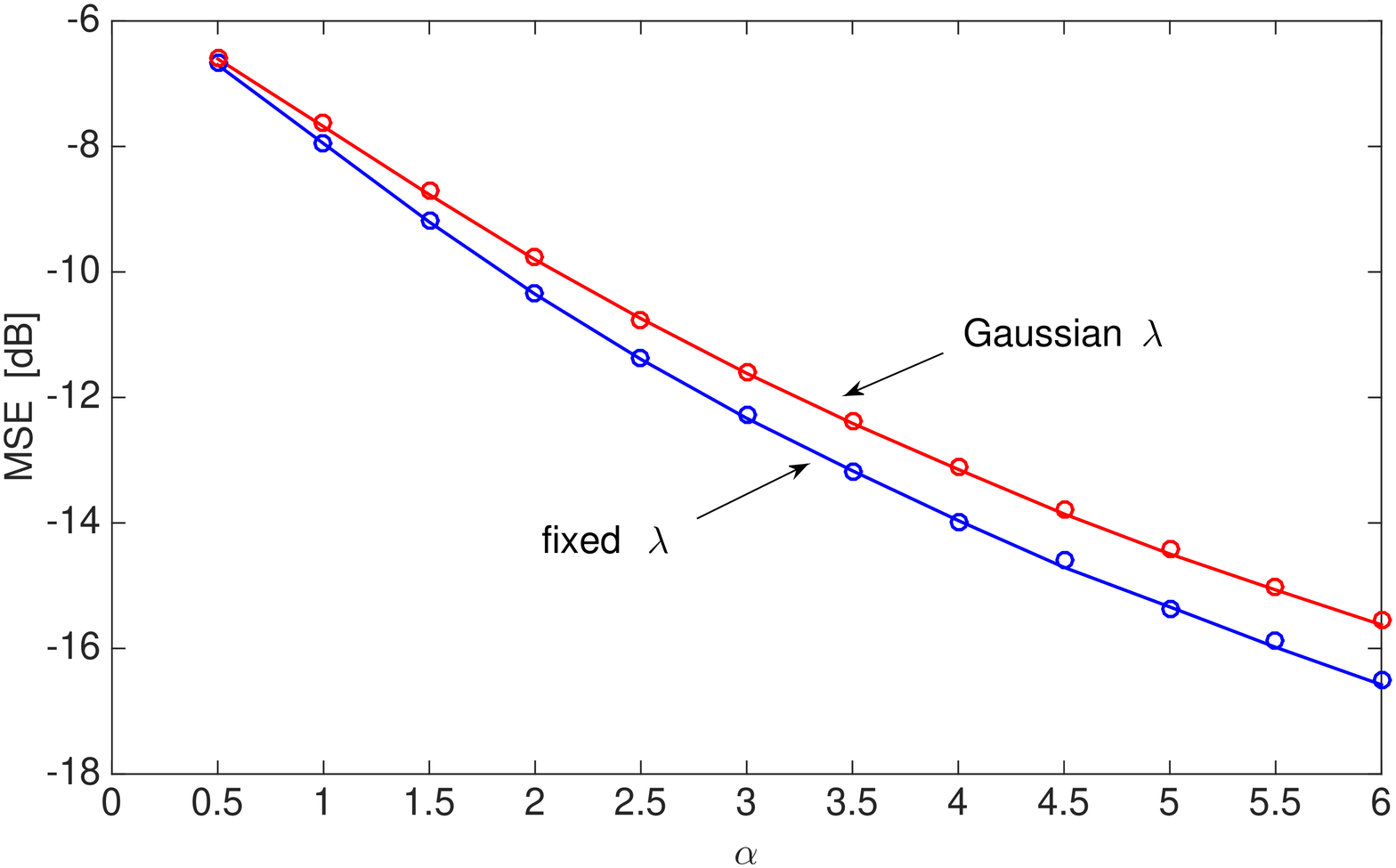}
  \caption{Lowest MSE [dB] (envelop) for each ratio $\alpha$ of signal $\rho = 0.25$, $\sigma_0^2 = 1$. The blue and red curves represent threshold strategies 1 and 2, respectively. The circles stand for the experimental estimate obtained using the CVX algorithm \cite{cvx} averaged over $1,000$ experiments with signal size $N = 128$ for each parameter set. }
  \label{fig:fixedVSGaussian}
\end{center}
\end{figure}

\begin{figure}[h]
\begin{center}
  \includegraphics[width=3.5in]{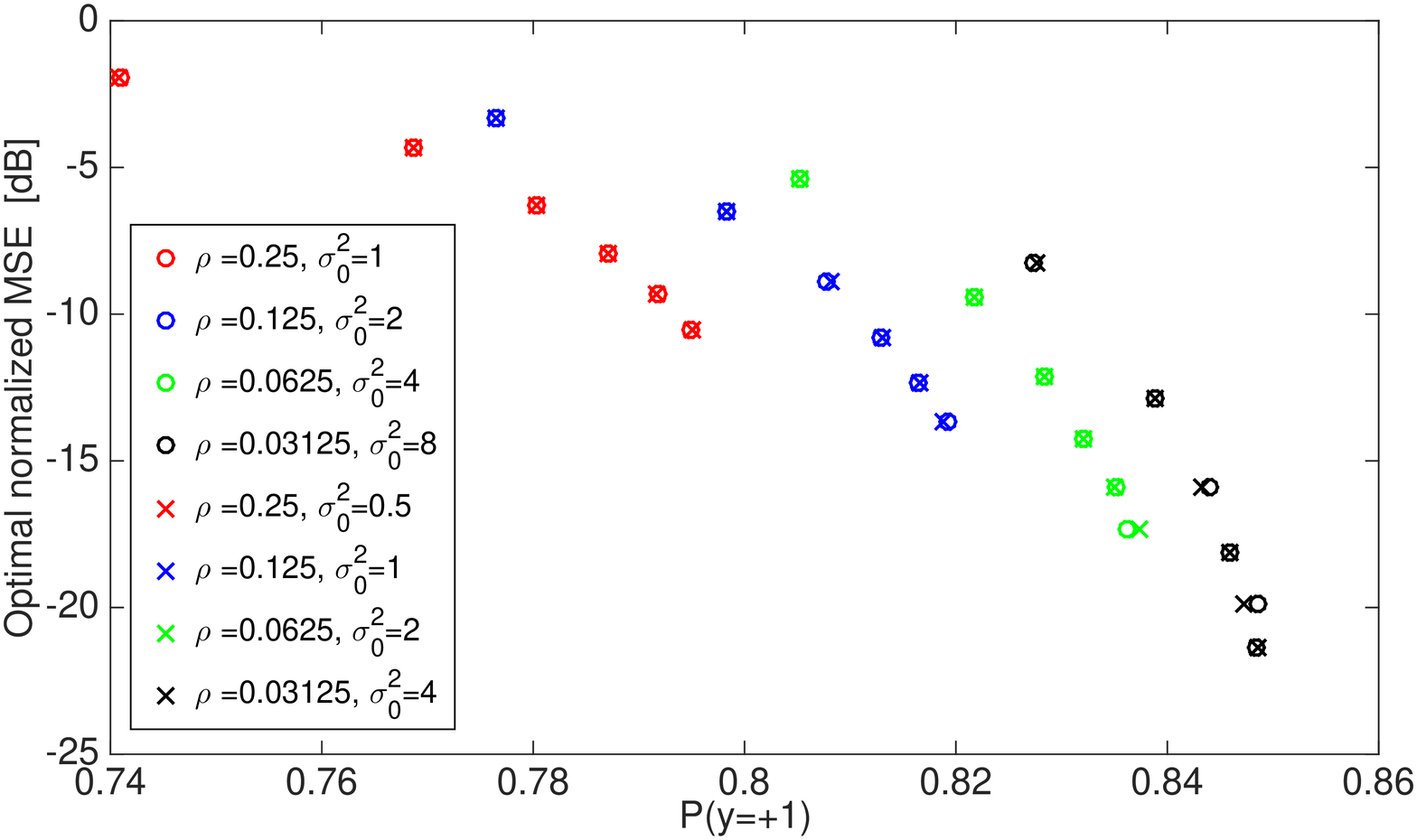}
  \caption{Optimal MSE: $\text{MSE}/(\rho\sigma_{0}^{2})$ in decibel versus the probability of $+1$ in $\vm{y}$ for fixed threshold 1-bit CS model. {{The plots are obtained by tuning $\lambda$ so as to minimize MSE and evaluating corresponding $P(y=1)$ from (\ref{eq:Py}) for each set of ($\rho, \sigma_{0}^{2}, \alpha$).}} Different colors represent varying signal sparsity; {red, blue, green and black mean $\rho=0.25, 0.125, 0.0625$ and $0.03125$, respectively. In each color, there are 6 circle and 6 cross symbols. Each of 6 symbols from left to right, corresponds to the result of $\alpha = 1, 2, \dots, 6$, respectively. Circles and crosses represent the results for $\rho\sigma_{0}^{2}=0.25$ and $0.125$, respectively, where $\rho\sigma_{0}^{2}$ means the ``power''(per component) of the original signal.}}
  \label{fig:normalizedMSE_Py}
\end{center}
\end{figure}

We solved the saddle point equations for signal sparsity $\rho = 0.25$ and variance $\sigma_{0}^{2} = 1$ when ratio $\alpha = 3$. The curve in Fig. \ref{fig:MSEdb_fix} denotes the theoretical predictions of MSE as evaluated by (\ref{qh_fix})--(\ref{m_fix}) (strategy 1) and (\ref{MSE}) plotted against the threshold $\lambda$. Fig. \ref{fig:MSEdb_Gauss} represents the theoretical predictions of MSE evaluated by (\ref{qh_Gauss})--(\ref{mh_Gauss}), (\ref{q_fix})--(\ref{m_fix}) (strategy 2), and (\ref{MSE}) plotted against the standard deviation $\sigma_{\lambda}$ of the threshold. Figures \ref{fig:MSEdb_fix} and \ref{fig:MSEdb_Gauss} show that there is an optimal threshold distribution (red circle symbol) that minimizes MSE for each set of parameters. Similar features hold for various sets of values of $\alpha, \rho, \sigma_{0}^{2}$ for both strategies 1 and 2. 

To compare the optimal MSE (changing threshold distribution) of strategy 1 and strategy 2, we plot the optimal MSE for the same signal distribution in Fig. \ref{fig:MSEdb_fix} and Fig. \ref{fig:MSEdb_Gauss} against $\alpha$ in Fig. \ref{fig:fixedVSGaussian}, which is referred to the envelope curve of MSE. The blue and red curves represent the envelope curves for strategies 1 and 2, respectively. From Fig. \ref{fig:fixedVSGaussian}, we can see that strategy 1 outperforms strategy 2 when parameters are optimally tuned. Therefore, we hereafter focus on strategy 1, for which the thresholds are fixed.  

The optimal value of $\lambda$ depends on $\rho$ and $\sigma_0^2$, which are not necessarily available in practice. To cope with such situations, we focus here on the distribution of binary output y, which indirectly conveys the information of $\rho\sigma_0^2$ and can be estimated from measurements. Fig. \ref{fig:normalizedMSE_Py} shows the relation between the optimal MSE and $P(y = +1)$ for eight signal distributions. For given $\lambda$, the probability of positive output $y = +1$ is evaluated as
\begin{eqnarray}
P(y=+1)&=&\frac{1}{M}\left[\prod_{\mu=1}^{M}\Theta\left(\vm{\Phi x_0}+\vm{\lambda}\right)_{\mu} \right]_{\vm{\Phi}, \vm{x_0}}\cr
&=&\left[\Theta\left(\vm{\Phi x_0}+\vm{\lambda}\right)_{\mu} \right]_{\vm{\Phi}, \vm{x_0}}\cr
&=&\int_{-\infty}^{\infty} \textrm{D}t \Theta\left(\sqrt{\rho}\sigma_{0}t+\lambda\right)\cr
&=&{H}\left(-\frac{\lambda}{\sqrt{\rho}\sigma_0}\right).
\label{eq:Py}
\end{eqnarray}
The horizontal axis in Fig. \ref{fig:normalizedMSE_Py} is calculated from (\ref{eq:Py}) by inserting the optimal value of $\lambda$. 
{{{The results indicate that when the signal is sparser (from red to black), corresponding $P(y = +1)$ is greater. Also, the value of $P(y = +1)$ that yields the optimal MSE monotonically increases with $\alpha$ when the signal distribution is fixed.}}
MSE is normalized by $\rho\sigma_0^2$ in order to eliminate its dependence on the scale of the original signal. 
From these results, we can see that the normalized MSE, $\text{MSE}/(\rho\sigma_{0}^{2})$, is the same when signal sparsity is the same. 
Although the optimal MSE depends on all system parameters $\rho$, $\sigma_0^2$, and compression rate $\alpha$, we can see that the corresponding $P(y = +1)$ is always placed in the range of $0.75 \sim 0.85$ for modest values of $1 \le \alpha \le 6$ in Fig. \ref{fig:normalizedMSE_Py}. In addition, the plots imply that although the optimal value of $P(y = +1)$ monotonically increases as $\alpha$ grows, it tends to converge to a value close to $0.85$.

\section{Learning algorithm for threshold}
The results of the last section suggest that for each parameter set, the optimal threshold that minimizes MSE is loosely characterized by the value of $P(y = +1)$, which can be statistically estimated from the outputs of measurements. This property may be utilized to adaptively tune the threshold for each measurement based on the results of previous measurements. 

A few studies have been conducted in the past on adaptive tuning of the threshold to improve signal reconstruction performance. For example, in \cite{adaptiveBaye}, given past measurements, a threshold value was determined to partition the consistent region along its centroid computed by generalized approximate message passing \cite{GAMP, 1bitCSBayeXuKaba}. However, in many realistic situations, precise knowledge of the prior distribution is unavailable, even if we might reasonably expect the signal to be sparse. Therefore, we will here develop a learning algorithm that can be executed without knowledge of the prior distribution of the signal. There is another general adaptive algorithm called $\Sigma\Delta$ quantization \cite{sigmadelta}. However, its goal is to find a satisfactory quantized representation of real number measurement and requires preprocessing based on real number measurements. Instead, the algorithm we develop aims to directly minimize MSE, and needs no preprocessing.

\begin{figure}
\renewcommand{\thepseudocode}{\arabic{pseudocode}}
\setcounter{pseudocode}{0}
\begin{pseudocode}[ruled]{adaptive thresholding}{\gamma, \delta, \lambda_{0}, U_0, V_0}

1)\ \mbox{\bf Initialization}:\\
 \hspace{15pt}\lambda \text{ seed}: \hspace{65pt}\lambda_0 \\
 \hspace{15pt}U \text{seed}: \hspace{65pt} U_{0}\GETS0 \\
 \hspace{15pt}V \text{seed}: \hspace{65pt} V_{0}\GETS0\\
\hspace{15pt}\text{Counter}: \hspace{62pt} k\GETS 0\\

2)\ \mbox{\bf Counter increase}:\\
 \hspace{20pt}k \GETS k+1\\

3)\ \mbox{\bf Measurement of signal}:\\
 \hspace{20pt} 
 y_{k}  = \textrm{sign}\left(\sum_{i} \vm{\Phi}_{k i} \vm{x_0}_{i} + \lambda_{k} \right)\\
 
4)\ \mbox{\bf Update $T_k$}:\\
 \hspace{20pt}
 U_k \GETS (y_k > 0) + \gamma U_{k - 1}\\
  \hspace{20pt}
 V_k \GETS 1 +\gamma V_{k - 1}\\
  \hspace{20pt}
 T_k \GETS U_{k}/V_{k}\\ 
  
5)\ \mbox{\bf Update $\lambda$}:\\ 
 \hspace{20pt}
 \lambda_k \GETS \lambda_{k - 1} + \delta\textrm{sign}(T - T_k)\\
 
6)\ \bold{Iteration}: \mbox{Repeat from 2) until $k = M$.}
%
\end{pseudocode}
\protect
\caption{\protect\label{algorithm}Pseudocode for adaptive thresholding of 1-bit CS measurements. Here, $y_k$ and $\lambda_k$ for $k = 1, 2, ..., M$ represent each element of vector $\vm{y}$ and $\vm{\lambda}$, respectively. Signal reconstruction can be carried out by versatile convex optimization algorithms. 
}
\label{algorithm}
\end{figure}

As shown in Fig. \ref{fig:normalizedMSE_Py}, MSE is minimized when $P(y = +1)$ takes a value of $0.75 \sim 0.85$ for various sets of parameters. To incorporate this property, we propose a strategy that first fixes a target value of $T$ for $P(y = +1)$, and tunes $\lambda$ so that an empirical distribution of $P(y = +1)$ approaches $T$. As we see in Fig. \ref{fig:normalizedMSE_Py}, for larger values of $\alpha$ or sparser signals, we should set $T$ as greater in the relevant range. There are various ways of estimating $P(y = +1)$ from the results of measurements. Of these, we use the damped average
\begin{equation}
T_\mu = \frac{\sum_{n = 0}^{\mu - 1}\gamma^{n}\delta_{y_{\mu -n}, +1}}{\sum_{n = 0}^{k - 1}\gamma^{n}}, 
\end{equation}
since it can be computed in an online manner as
\begin{equation}
T_{\mu +1} = \frac{\gamma \left (1 - \gamma^\mu \right )T_\mu + (1 - \gamma) \delta_{y_{\mu + 1}, +1}}{1 - \gamma^{\mu + 1}}, 
\end{equation}
which does not require referring to the details of previous measurements. Here, the damping factor $\gamma$ is a parameter that we have to set. In experiments, we set $\gamma = 0.8$; but as long as we tested it, the obtained performance was not particularly sensitive to the choice of this parameter. (\ref{measurement}) indicates that $P(y = +1)$ monotonically increases as $\lambda_\mu$ grows. This implies that $\lambda_\mu$ should be increased when $T > T_{\mu - 1}$, and decreased otherwise. To implement this idea, we design the learning algorithm of $\lambda_\mu$ as

\begin{equation}
\lambda_\mu = \lambda_{\mu - 1} + \delta \textrm{sign}(T - T_{\mu - 1}), 
\end{equation}
where $\delta$ denotes the step size that is also set by users. The pseudocode for adaptive thresholding 1-bit CS measurements is shown in Fig. \ref{algorithm}. Following measurement, signal reconstruction can be carried out by versatile convex optimization algorithms \cite{cvx} by solving (\ref{thresh1bitCS}).

\begin{figure}[t]
\begin{center}
  \includegraphics[width=3.5in]{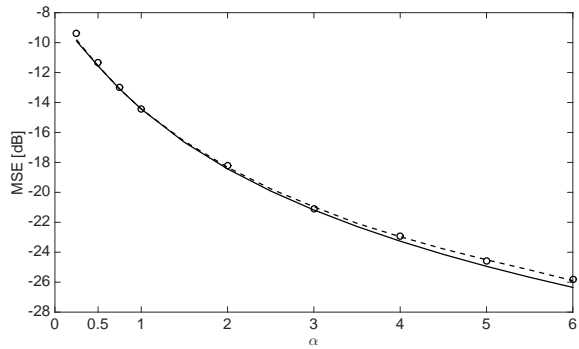}
  \caption{Experimental result from the adaptive thresholding algorithm for signal $\rho = 0.0625$, $\sigma_{0}^2 = 2$, and $N = 128$. The circles denote the average of $1,000$ experiments. The parameter settings were $T = 0.8$, $\gamma = 0.8$, $\lambda_0 = 0.5$, and $\delta = 0.01$. The broken line represents the replica prediction when $\lambda$ is set to offer $P(y = +1) = 0.8$ while the full curve denotes this for optimally tuned $\lambda$.}
  \label{fig:cvxresult}
\end{center}
\end{figure}

Since we plan to apply the adaptive algorithm in situations involving a finite number of measurements, the extent to which the initial threshold $\lambda_0$ is remote from the optimal threshold $\lambda_{\text{opt}}$, which is unknown beforehand, and the variation in the step size $\delta$ may significantly influence reconstruction performance. In order to set an appropriate value of $\lambda_0$, we propose testing it by measuring the signal a few times. If the outputs are limited almost exclusively to $+1$ or $-1$, we change the threshold through the bisection method, which involves dividing or multiplying it by $2$ until the outputs are adequately mixed with $+1$ and $-1$. The resulting threshold should yield an appropriate value of $\lambda_0$ close to $\lambda_{\text{opt}}$. Having set $\lambda_0$, an appropriate value of $\delta$ should be in smaller order in order to tweak it to $\lambda_{\text{opt}}$.

The results of our numerical experiments are shown in Fig. \ref{fig:cvxresult} as circles. Each circle denotes the average of $1,000$ experiments for systems where $N = 128$. The parameter settings of the experiments were $T = 0.8$, $\gamma = 0.8$, $\lambda_0 = 0.5$, and $\delta = 0.01$ for signal distribution $\rho = 0.0625$ and $\sigma_{0}^2 = 2$. The solid line in Fig. \ref{fig:cvxresult} represents the envelop for MSE (dB) for each $\alpha$. On the other hand, the dashed curve represents the prediction of MSE (dB) using replica analysis when $P(y = +1) = 0.8$, which was achieved by $\lambda = 0.2976$ according to (\ref{eq:Py}). Fig.\ref{fig:cvxresult} shows that the adaptive thresholding algorithm in conjunction with the employment of CVX for signal reconstruction can achieve nearly the same performance in terms of MSE as the statistical prediction for $P(y = +1) = 0.8$, and the result is reasonably close to the envelope MSE.

\section{Conclusion}
{{In this paper, we analyzed the typical performance of the thresholding 1-bit compressed sensing, which can reconstruct both the scaling and the directional information of the signal.}}
Considering the most general situation, where no detailed prior knowledge of sparse signals is available, we employed the $l_1$-norm minimization approach. By utilizing the replica method from statistical mechanics, the mean squared error behavior of reconstruction for standard i.i.d measurement matrix and i.i.d Bernoulli-Gaussian signal was derived in the large system size limit. We compared two design strategies for the elements of the threshold vector, which corresponded to setting a fixed or random value as threshold. Our analysis showed that the fixed threshold strategy can achieve lower MSE than the random threshold strategy statistically.

Another observation from the replica results was that there is an optimal threshold that minimizes MSE for a set of signal distributions and measurement ratios. However, in order to evaluate the optimal threshold, we need to know the prior distribution of the signal, which is not necessarily available in practical situations. Therefore, we shifted our focus to the relation between the optimal threshold and the distribution of the binary outputs, which can be empirically evaluated from signal measurements. The replica analysis indicated that the MSE is minimized when $P(y = +1)$ is set in the vicinity of $0.75 \sim 0.85$ for a wide region of system parameters.

On the basis of this observation, an algorithm that adaptively tunes the threshold at each measurement in order to obtain $P(y = +1)$ close to our target value was proposed. Combined with versatile convex optimization algorithms, the adaptive thresholding algorithm offers a computationally feasible and widely applicable 1-bit CS scheme. Numerical experiments showed that it can yield nearly optimal performance, even when no detailed prior knowledge of sparse signals is available.

Improvements on the adaptive thresholding algorithm as well as the application of the algorithm to practical problems form part of our future research in the area.

\ack
YX is supported by JSPS Research Fellowships DC2. This study was partially supported by JSPS KAKENHI Nos. 26011287 (YX) and 25120013 (YK). 

\appendix
\section{Derivation of $(\ref{eq:free energy})$}
\label{replicaderivation}
\subsection{Assessment of $\left[Z^{n}\left(\beta; \vm{\Phi}, \vm{x}^{0},\vm{\lambda}\right)\right]_{\vm{\Phi},\vm{x}^{0},\vm{\lambda}}$ for $n \in \mathbb{N}$}
Averaging (\ref{eq:expansion}) with respect to $\vm{\Phi}$ and $\vm{x}^0$ offers the following expression of the $n$-th moment of the partition function:
\begin{eqnarray}
&&\left[Z^{n}\left(\beta; \vm{\Phi}, \vm{x}^{0},\vm{\lambda}\right)\right]_{\vm{\Phi},\vm{x}^{0},\vm{\lambda}}\cr
&=&
\int \prod_{a=1}^n \left (d \textrm{\boldmath $x$}^a e^{-\beta||\textrm{\boldmath $x^a$}||_{1}} \right ) 
\times 
\left[\prod_{a=1}^n \prod_{\mu=1}^M 
\Theta\left(
(\vm{\Phi} \vm{x}^0+\vm{\lambda})_\mu 
(\vm{\Phi} \vm{x}^a+\vm{\lambda})_\mu 
\right)
\right]_{\vm{\Phi},\vm{x}^{0},\vm{\lambda}}. 
\label{Zmoment}
\end{eqnarray}
We insert $n(n + 1)/2$ trivial identities
\begin{equation}
1 = N \int dq_{ab} \delta \left (\vm{x}^a \cdot \vm{x}^b - N q_{ab} \right ), 
\end{equation}
where $a > b = 0, 1, 2, \ldots, n$, into (\ref{Zmoment}). Furthermore, we define a joint distribution of $n + 1$ vectors $\{\vm{x}^a\} = \{\vm{x}^0, \vm{x}^1, \vm{x}^2, \ldots, \vm{x}^n \}$ as
\begin{equation}
P\left (\{\vm{x}^a\} |\vm{Q}\right )= \frac{1}{V\left (\vm{Q}\right ) } P(\vm{x}^0) \times 
\prod_{a=1}^n \left (
e^{-\beta||\textrm{\boldmath $x^a$}||_{1}} \right )  
\times \prod_{a>b} \delta \left (\vm{x}^a \cdot \vm{x}^b - N q_{ab} \right ), 
\label{replica_joint_dist}
\end{equation}
where $\vm{Q} = (q_{ab})$ is an $(n + 1) \times (n + 1)$ symmetric matrix whose $00$ and the other diagonal entries are fixed as $\rho$ and $q_{aa}$, respectively. $P(\vm{x}^0) = \prod_{i = 1}^N \left ((1 - \rho)\delta(x_i^0) + \rho \tilde{P}(x_i^0) \right )$ denotes the distribution of the original signal $\vm{x}^0$, and $V\left (\vm{Q}\right )$ is the normalization constant that makes $\int \prod_{a = 0}^n d \vm{x}^a P\left (\{\vm{x}^a\} |\vm{Q}\right ) = 1$ hold. These indicate that (\ref{Zmoment}) can also be expressed as
\begin{eqnarray}
\left[Z^{n}\left(\beta; \vm{\Phi}, \vm{x}^{0},\vm{\lambda}\right)\right]_{\vm{\Phi},\vm{x}^{0},\vm{\lambda}}
=\int d\vm{Q} \left (V\left (\vm{Q}\right ) \times \left[\Xi\left (\vm{Q}\right ) \right]_{\vm{\lambda}}\right ), 
\label{Zn}
\end{eqnarray}
where $d\vm{Q} \equiv \prod_{a>b}d q_{ab}$ and 
\begin{equation}
\Xi\left (\vm{Q}\right )
\!=\!\int \prod_{a=0}^n d\vm{x}^a P\left (\{\vm{x}^a\} |\vm{Q} \right ) 
\times\left[\prod_{a=1}^n \prod_{\mu=1}^M 
\Theta\left(
(\vm{\Phi} \vm{x}^0+\vm{\lambda})_\mu 
(\vm{\Phi} \vm{x}^a+\vm{\lambda})_\mu\! 
\right)\!
\right]_{\vm{\Phi}}\!. 
\label{Xi}
\end{equation}

Equation (\ref{Xi}) can be regarded as the average of $\prod_{a = 1}^n \prod_{\mu = 1}^M \Theta\left((\vm{\Phi} \vm{x}^0 + \vm{\lambda})_\mu (\vm{\Phi} \vm{x}^a + \vm{\lambda})_\mu \right)$ with respect to $\{\vm{x}^a\}$ and $\vm{\Phi}$ over distributions of $P\left (\{\vm{x}^a\} \right )$ and $P(\vm{\Phi}) \equiv \left (\sqrt{2\pi/N} \right )^{-MN} \exp \left (-(N/2) \sum_{\mu, i} \Phi_{\mu i}^2 \right )$. In computing this, it is noteworthy that when $N$ and $M$ tend to infinity while keeping $\alpha = \frac{M}{N}$ finite, the Central Limit Theorem guarantees that $u_\mu^a \equiv (\vm{\Phi} \vm{x}^a)_\mu = \sum_{i = 1}^N \Phi_{\mu i} x_i^a$ can be handled as zero-mean multivariate Gaussian random numbers whose variance and covariance are provided by 
\begin{equation}
\left [u_\mu^a u_\nu^b \right ]_{\vm{\Phi},\{\vm{x}^a\}}
=\delta_{\mu \nu} q_{ab}, 
\end{equation}
when $\vm{\Phi}$ and $\{\vm{x}^a\}$ are generated independently from $P(\vm{\Phi})$ and $P\left (\{\vm{x}^a\} \right )$, respectively. This means that (\ref{Xi}) can be evaluated as
\begin{equation}
\Xi(\vm{Q}) =
\!\left (\!\frac{\!\int\! d \vm{u} \exp\left (-\frac{1}{2} \vm{u}^{\rm T} \vm{Q}^{-1} \vm{u} \right )
\prod\limits_{a=1}^{n}\! \Theta\!\left (\!(u^0 \!+\!\lambda) (u^a \!+\!\lambda)\!\right )  }{
(2\pi)^{(n+1)/2} (\det \vm{Q})^{1/2}}\!\right )^{M}, 
\label{logXi}
\end{equation}
where $u^0$, $u^{a}$, and $\lambda$ represent the typical elements of $\vm{u^0}$, $\vm{u^{a}}$ and $\vm{\lambda}$, respectively, since each $\mu$ is independently distributed.

On the other hand, expression 
\begin{equation}
\delta\!\left (\!\vm{x}^a \cdot \vm{x}^b \!-\!N q_{ab}\right )\!
=\!\frac{1}{2 \pi} \int_{-{\rm i}\infty}^{+{\rm i}\infty}
d \hat{q}_{ab}  e^{\hat{q}_{ab} 
\left (\vm{x}^a \cdot \vm{x}^b-N q_{ab} \right )}, 
\end{equation}
and use of the saddle point method offer 
\begin{equation}
\frac{1}{N} \ln V(\vm{Q})
\!=\mathop{\rm extr}_{\hat{\vm{Q}}}
\!\left \{\!
-\!\frac{1}{2} {\rm Tr} \hat{\vm{Q}}\vm{Q} \right . 
\left . \!+ \!\ln 
\!\left (
\!\int\! d\vm{x} 
P(x^0) \exp\! \left (\!\frac{1}{2} 
\vm{x}^{\rm T} \hat{\vm{Q} }\vm{x}\!-\!\beta \sum_{a=1}^n \beta |x^a| \!\right )\!
\right )\! 
\right \}\!.  
\label{logV}
\end{equation}
Here, $\vm{x} = (x^0, x^1, \ldots, x^n)^{\rm T}$, and $x^a$ represents the typical element of $\vm{x}^a$, since each $x_{i}^{a}$ is independently distributed. $\hat{\vm{Q}}$ is an $(n + 1) \times (n + 1)$ symmetric matrix whose $00$ and other diagonal components are given as $0$ and $-\hat{q}_{aa}$, respectively, while off-diagonal entries are offered as $\hat{q}_{ab}$. Equations (\ref{logXi}) and (\ref{logV}) indicate that $N^{-1} \ln \left [Z^n (\beta; \vm{\Phi}, \vm{x}^0, \vm{\lambda} ) \right]_{\vm{\Phi}, \vm{x}^0, \vm{\lambda}}$ is correctly evaluated by using the saddle point method with respect to $\vm{Q}$ in the assessment of the right-hand side of (\ref{Zn}), when $N$ and $M$ tend to infinity while keeping $\alpha = M/N$ finite. 

\subsection{Treatment under the replica symmetric ansatz}
Let us assume that the relevant saddle point in assessing (\ref{Zn}) is of the form of (\ref{RSanzats}) and, accordingly, 
\begin{eqnarray}
\hat{q}_{ab}=\hat{q}_{ba} =\left \{
\begin{array}{ll}
0, &( \mbox{$a=b=0$})\\
\hat{m}, &( \mbox{$a = 1, 2, \ldots, n$; $b = 0$})\\
\hat{Q}, &( \mbox{$a = b = 1, 2, \ldots, n$})\\
\hat{q}, &( \mbox{$a \ne b = 1, 2, \ldots, n$})
\end{array}
\right .  .
\label{RShatQ}
\end{eqnarray}
The $n + 1$-dimensional Gaussian random variables $u^0, u^1, \ldots, u^n$, whose variance and covariance are provided as (\ref{RSanzats}), can be expressed as 
\begin{eqnarray}
&& u^0=\sqrt{\rho{\sigma_{0}^2}-\frac{m^2}{q}}s^0 + \frac{m}{\sqrt{q}} t, \label{Newgauss0}\\
&& u^a=\sqrt{Q-q} s^a+ \sqrt{q} t, \ (a=1,2,\ldots,n) \label{Newgaussa} 
\end{eqnarray}
by utilizing $n + 2$ independent standard Gaussian random variables $t$ and $s^0, s^1, \ldots, s^n$. This indicates that (\ref{logXi}) is evaluated as
\begin{eqnarray}
\Xi(\vm{Q})=\left (
 \int {\rm D}t 
{H}\left (-\frac{\frac{m}{\sqrt{q}}t+\lambda}{\sqrt{\rho{\sigma_{0}^2} -\frac{m^2}{q}} } \right )
{H}^n \left (-\frac{\sqrt{q}t+\lambda }{\sqrt{Q-q}}\right ) \right.\nonumber\\
\left.+{H}\left (\frac{\frac{m}{\sqrt{q}}t+\lambda}{\sqrt{\rho{\sigma_{0}^2} -\frac{m^2}{q}} } \right )
{H}^n \left (\frac{\sqrt{q}t+\lambda }{\sqrt{Q-q}}\right ) 
\right )^M. 
\label{NewXi}
\end{eqnarray}
On the other hand, substituting (\ref{RShatQ}) into (\ref{logV}), in conjunction with the identity,
\begin{equation}
\exp \!\left (\!\hat{q} \!\sum_{a>b(\ge 1)}\! x^a x^b \!\right )
\!=\!\int \!{\rm D}z \exp \left (\sum_{a=1}^n 
\left (
-\frac{\hat{q}}{2} (x^a)^2 \!+\! \sqrt{\hat{q}} z x^a \!\right ) \!\right ) \!
\end{equation}
where $z$ is a standard Gaussian random variable, yields
\begin{eqnarray}
&&\frac{1}{N} \ln V(\vm{Q} )=\mathop{\rm extr}_{\hat{Q}, \hat{q},\hat{m} }
\left \{
\frac{n}{2}\hat{Q}Q-\frac{n(n-1)}{2} \hat{q}q -\hat{m} m{\sigma_{0}^2} \right . \cr
&&\left . 
+\ln 
\left [\left (
\!
\int 
\!
dx \exp 
\!\left (
\!
-\frac{\hat{Q}
\!+\! \hat{q}}{2} x^2\!+\!\left (\!\sqrt{\hat{q}}z \!+\! \hat{m} x^0 \!\right )
\!x \! \right.\right.\right.\right.
\left.\left.\left.\left.
-\!\beta |x| \right )
\right )^n \right ]_{x^0,z} \right \}. 
\label{NewV}
\end{eqnarray}
Although we have assumed that $n \in \mathbb{N}$, the expressions of (\ref{NewXi}) and (\ref{NewV}) are likely to hold for $n \in \mathbb{R}$ as well. Therefore the average free energy $\overline{f}$ can be evaluated by substituting these expressions into the formula $\overline{f}= -\lim_{n \to 0} (\partial/\partial n) \left ((\beta N)^{-1} \ln \left [Z^n (\beta; \vm{\Phi}, \vm{x}^0 , \vm{\lambda}) \right]_{\vm{\Phi}, \vm{x}^0}, \vm{\lambda}\right )$. 

In the limit of $\beta \to \infty$, a nontrivial saddle point is obtained only when $\chi \equiv \beta (Q - q)$ is kept finite. Accordingly, we change the notations of the auxiliary variables as $\hat{Q} + \hat{q} \to \beta \hat{Q}$, $\hat{q} \to \beta^2 \hat{q}$, and $\hat{m} \to \beta \hat{m}$. Furthermore, we use the asymptotic forms
\begin{eqnarray}
\lim_{\beta \to \infty}\frac{1}{\beta}\int {\rm D}t
{H}\left (\frac{\frac{m}{\sqrt{q}}t+\lambda}{\sqrt{\rho{\sigma_{0}^2} -\frac{m^2}{q}} }\right )
\ln {H}\left (\frac{\sqrt{q}t+\lambda }{\sqrt{Q-q}}\right ) \cr
\!=\!\int{\rm D}t
{H}\!\left (\!\frac{\frac{m}{\sqrt{q}}t+\lambda}{\sqrt{\rho{\sigma_{0}^2} -\frac{m^2}{q}} }\! \right )\!\left (\!-\!\frac{(\sqrt{q}t+\lambda)^2}{2\chi} \Theta(\sqrt{q}t+\lambda)\! \right ) \!
\end{eqnarray}
and 
\begin{eqnarray}
&&\lim_{\beta \to \infty}\frac{1}{\beta}
\ln 
\left (
\!
\int 
\!
dx \exp 
\!\left (\beta \left (
\!
-\frac{\hat{Q}}{2} x^2\!+\!\left (\!\sqrt{\hat{q}}z \!+\! \hat{m} x^0 \!\right )
\!x \!-\! |x| \right ) \right )
\right ) \cr
&& =-\phi\left (\sqrt{\hat{q}}z+\hat{m}x^0;\hat{Q} \right ). 
\end{eqnarray}
Using these in the resultant expression of $\overline{f}$ offers (\ref{eq:free energy}).



\section*{References}
\begin{thebibliography}{99}
\bibitem{CScandes}
Candes E 2006 {\it Proc. Int. Congr. Math. (Madrid)}
\bibitem{CSdonoho}
Donoho D 2006 {\it IEEE Tans. Inf. Theory} {\bf 6} 4 p~ 1289-1306
\bibitem{Elad2010}
Elad M 2010 {\it Sparse and Redundant Representations: From Theory to Applications in Signal and Image Processing}
New York: Springer
\bibitem{StarckMurtaghFadili2010}
Starck J, Murtagh F and Fadili J M 2010 {\it Sparse Image and Signal Processing: Wavelets, Curvelets, Morphological Diversity}
(New York: Cambridge University Press)
\bibitem{CandesWakin2008}
Cand\`{e}s E J and Wakin M B 2008 {\it IEEE Signal Processing Magazine} {\bf 21}
\bibitem{CandesRombergTao2006}
Cand\`{e}s E J, Romberg J and Tao T 2006  {\it IEEE Trans. Inform. Theory} {\bf 52} 489.
\bibitem{KWT2009}
Kabashima Y, Wadayama T and Tanaka T 2009 
{\it J. Stat. Mech.} L09003; 2012 {\it J. Stat. Mech.} E07001
\bibitem{Sompolinsky2010}
Ganguli S and Sompolinsky H 2010 {\it Phys. Rev. Lett.} {\bf 104} 188701
\bibitem{Krzakala2012}
Krzakala F, M\'{e}zard M, Sausset F, Sun Y F and Zdeborov\'{a} L 2012 {\it Phys. Rev. X} {\bf 2} 021005.
\bibitem{bandlimit}
Tropp J, Laska J, Duarte M, Romberg J and Baranuik R 2010
{\it IEEE Trans. Inf. Theory}
{\bf 56} 1 p~520--44
\bibitem{singlepixel}
Duarte M, Davenport M, Takhar D, Laska J, Sun T, Kelly K and Baranuik R 2008
{\it IEEE Signal Process. Mag.}{\bf 25} 2 p~83--91
\bibitem{measurementsVSbits}
Sarvotham S, Baron D and Baranuik R 2006 {\it Proc. of 44th Allerton Conf. Comm., Ctrl., Computing.}
\bibitem{ratedistortion}
Fletcher A K, Rangan S and Goyal V K 2007 {\it Proc. of IEEE Inter. Conf. on Acoustics, Speech and Signal Processing}(Honolulu) {\bf 3}
\bibitem{1bitCSbaranuik}
Boufounos P and Baranuik R, 2008 {\it Proc. 42nd Annu. Conf. Inf. Sci. Syst.} (Princeton, NJ) p~ 16--21
\bibitem{1bitCSL1XuKaba}
Xu Y and Kabashima Y 2013 {\it J. Stat. Mech.} P02041
\bibitem{1bitCSNormEstKKR}
Knudson K, Saab R and Ward R 2014 {\it One-bit compressive sensing with norm estimation}  arXiv: 1404.6853
\bibitem{cvx}
Grant M, Boyd S and Ye Y, {\it CVX: Matlab Software for Disciplined Convex Programming} [Online] Available: \texttt{http://cvxr.com/cvx/}
\bibitem{replica} 
Dotsenko V S 2001 {\it Introduction to the replica theory of disordered statistical systems} Cambridge: Cambridge University Press.
\bibitem{adaptiveBaye}
Kamilov U S, Bourquard A, Amini A and Unser M 2012 {\it IEEE Signal Process. Letters} {\bf 19} 10 p~607--610
\bibitem{GAMP}
Rangan S 2010 {\it Proc. IEEE Int. symp. on Information Theory} (St. Petersburg, Russia p~2168--72
\bibitem{1bitCSBayeXuKaba}
Xu Y and Kabashima Y and Zdeborova L 2014 {\it J. Stat. Mech.} P11015
\bibitem{sigmadelta}
Boser B and Wooley B 1988 {\it IEEE Journal of Solid-State Circuits} {\bf 23}6  p~1298--1308
\end{thebibliography}
\end{document}